\documentclass[letter,oldversion]{aa}
\usepackage{graphicx}
\usepackage{txfonts}
\usepackage{natbib}
\bibpunct{ (}{)}{;}{a}{}{,}

\begin{document}

\title{Calibration of the galaxy cluster $M_{\rm 500}$--$Y_{\rm X}$
  relation with XMM-Newton} 
\author{M. Arnaud \inst{1},
              E. Pointecouteau \inst{2} and
                      G.W. Pratt \inst{3}}
\offprints{M. Arnaud, \email{Monique.Arnaud@cea.fr}}

 \institute{
 $^1$ Laboratoire AIM, DAPNIA/Service d'Astrophysique - CEA/DSM - CNRS
 - Universit\'{e} Paris Diderot, B\^{a}t. 709, CEA-Saclay, F-91191
 Gif-sur- Yvette Cedex, France \\ 
 $^2$ CESR, 9 Av du colonel Roche, BP 44346, 31028 Toulouse Cedex 4,
 France\\ 
 $^3$ Max-Planck-Institut f\"ur extraterrestriche Physik, Giessenbachstra{\ss}e, 85748 Garching, Germany}

  \date{Received 24 August 2007; accepted 9 September 2007}
  \abstract 
   {The quantity $Y_{\rm X}$, the product of the
     X-ray temperature $T_{\rm X}$ and gas mass $M_{\rm g}$, has
     recently been proposed as a robust low-scatter mass indicator for
     galaxy clusters. 
    Using precise measurements from XMM-Newton data of a sample of 10
    relaxed nearby clusters, spanning a $Y_{\rm X}$ range of
     $10^{13}$--$10^{15}$\,M$_\odot$\,keV, we investigate  the
     $M_{500}$--$Y_{\rm X}$ relation.
     The $M_{500}$ -- $Y_{\rm
       X}$ data exhibit a power law relation with slope
     $\alpha=0.548 \pm 0.027$, close to the self-similar value (3/5) and
     independent of the mass range considered. However, the 
     normalisation is $\sim 20\%$  below the prediction from numerical
     simulations including cooling and galaxy feedback.  We discuss two effects that could contribute to the normalisation offset: an
     underestimate of the true mass due to  
     the HE assumption used in X-ray mass estimates, and an underestimate of
     the hot gas mass fraction in the simulations. A comparison of
     the functional form and scatter of the relations between
     various observables and the mass suggest that $Y_{\rm X}$
     may indeed be a better mass proxy than $T_{\rm X}$ or $M_{\rm
       g,500}$.}  

   \maketitle
%
%________________________________________________________________
\def\etal{et al.}

\def\Mgv{M_{\rm g,500}}
\def\Mg{M_{\rm g}}
\def\YX {Y_{\rm X}}
\def\TX {T_{\rm X}}
\def\fgv {f_{\rm g,500}}
\def\fg  {f_{\rm g}}
\def\kT {{\rm k}T}
\def\Mv {M_{\rm 500}}
\def \Rv {R_{500}}
\def\keV {\rm keV}

\def\MT {$M_{500}$--$T_{\rm X}$}
\def\MY {$M_{500}$--$Y_{\rm X}$}
\def\MMg {$M_{500}$--$M_{\rm g,500}$}
\def\MgT {$M_{\rm g,500}$--$T_{\rm X}$}
\def\MgY {$M_{\rm g,500}$--$Y_{\rm X}$}

\def\msol {{\rm M_{\odot}}}

\def\lesssim{\mathrel{\hbox{\rlap{\hbox{\lower4pt\hbox{$\sim$}}}\hbox{$<$}}}}
\def\gtrsim{\mathrel{\hbox{\rlap{\hbox{\lower4pt\hbox{$\sim$}}}\hbox{$>$}}}}

% satellites
\def \xmm {\hbox{\it XMM-Newton}}
\def \chandra {\hbox{\it Chandra }}

\section{Introduction}

All theoretical approaches characterise galaxy clusters in terms of
their mass.  Models  of
structure formation predict the space density, distribution
and physical properties of clusters as a function of mass and redshift
\citep[e.g. ][]{bertschinger98}. However, the mass is not easily measured.  
X-ray estimates from the hydrostatic equilibrium (HE) equation are
valid only for reasonably relaxed clusters and require temperature
profiles of high statistical quality; furthermore, the available
precision rapidly degrades with redshift.  
Based on the regularity of the cluster population, other X-ray
observables such as luminosity, temperature $\TX$, or gas mass $\Mg$,
have been used as proxies for the mass, e.g. to constrain cosmological
parameters using cluster surveys \citep{voi05}.  Studies of cluster
formation physics must also rely on mass proxies when considering
unbiased (i.e., covering a variety of dynamical 
states) or distant cluster samples \citep[e.g.][]{mau07}.  
The identification of the best mass proxy, and knowledge
of its exact relation to the mass, are therefore important.

The most commonly used mass proxy, $\TX$, is expected to
be closely related to the mass via the virial theorem.  
Significant progress on the calibration of the local $M$--$\TX$
relation for {\it relaxed} clusters has recently been made, with
excellent agreement now achieved between various observations
\citep{app05,vik06}, and comparison between observations and numerical
models including cooling and galaxy feedback showing agreement to the
$\sim10\%$  level \citep[e.g.][]{nag07b,app05}.  \citet{kra06}
recently proposed a new mass proxy,  $\YX=\TX\Mgv$, where
$\Mgv$ is the gas mass within $\Rv$, the radius corresponding to a
density contrast of  $\delta=500$. $\YX$ is related to the thermal
energy of the gas and is the X-ray analogue of the integrated SZ
Comptonisation parameter, $Y_{\rm SZ}$. The numerical simulations of \citeauthor{kra06} showed
that, as compared to $\TX$ or $\Mgv$,  $\YX$ is a better mass proxy,
in the sense that the intrinsic scatter was lower than for any other
mass indicator, regardless of cluster dynamical state (similar to
previous results for the $M$--$Y_{\rm SZ}$ relation, e.g.,  
\citealt{das04,mot05,nag06}). Furthermore, its evolution appears to be
close to the standard self-similar 
expectation.   

In this Letter, we present  the \MY\ relation derived from precise
\xmm\ data and compare it to the \MMg\ and \MT\ relations. The
\MY\  relation is discussed with respect to previous \chandra\ results
and theoretical expectations \citep{nag07b}.  Other relations between
observables,  such as the variation of the gas mass fraction $\fgv$
with mass,  are also investigated in order to shed new light on the
scatter and slope of the various mass-proxy relations.   

\begin{table*}
     \caption[]{Physical cluster parameters.  $\Mv$, $M_{\rm g, 500}$
       and  $f_{\rm g,500}$ are the total mass, gas mass and gas mass
       fraction respectively, within the radius $\Rv$, inside which
       the mean mass density is 500 times the critical
       density at the cluster redshift. $\TX$ is the
       spectroscopic temperature within $[0.15-0.75]R_{500}$ and
       $Y_{\rm X}= M_{\rm g, 500} \TX$. Values  are given for a
       $\Lambda$CDM cosmology  with $\Omega_{\rm m}=0.3$,
       $\Omega_{\Lambda}=0.7$, $H_0=70$~km s$^{-1}$ Mpc$^{-1}$. Errors
       are $1\,\sigma$.} 
     \label{tab:clu}
     \begin{center}
%    $$
    \begin{tabular}{lccccccccc}
    \hline
    \hline
    Cluster & $z$ & $\TX (\keV)$&  $M_{500}~(10^{14}\, \msol)$& $M_{\rm g,500}~(10^{13}\, \msol)$&$Y_{\rm X}~(10^{13}\, \msol\,\keV)$& $f_{\rm g,500}$ \\
    \hline
\object{A 1983}& $0.0442$ & $2.18\pm0.09$ & $1.09_{-0.29}^{+0.45}$ & $ 0.64_{- 0.08}^{+ 0.10}$ & $ 1.39_{-0.18}^{+0.23}$ & $0.058_{-0.017}^{+0.026}$\\
\object{MKW9}& $0.0382$ & $2.43\pm0.24$ & $0.88_{-0.18}^{+0.23}$ & $ 0.49_{- 0.05}^{+ 0.06}$ & $ 1.19_{-0.17}^{+0.18}$ & $0.055_{-0.012}^{+0.016}$\\
 \object{A 2717}& $0.0498$ & $2.56\pm0.06$ & $1.10_{-0.11}^{+0.13}$ & $ 1.02_{- 0.05}^{+ 0.06}$ & $ 2.60_{-0.15}^{+0.16}$ & $0.093_{-0.011}^{+0.012}$\\
\object{A 1991}& $0.0586$ & $2.71\pm0.07$ & $1.20_{-0.12}^{+0.13}$ & $ 1.25_{- 0.06}^{+ 0.06}$ & $ 3.39_{-0.19}^{+0.19}$ & $0.104_{-0.011}^{+0.012}$\\
 \object{A 2597}& $0.0852$ & $3.67\pm0.09$ & $2.22_{-0.21}^{+0.23}$ & $ 2.51_{- 0.08}^{+ 0.09}$ & $ 9.21_{-0.38}^{+0.39}$ & $0.113_{-0.011}^{+0.012}$\\
 \object{A 1068}& $0.1375$ & $4.67\pm0.11$ & $3.87_{-0.27}^{+0.29}$ & $ 3.77_{- 0.10}^{+ 0.10}$ & $17.6_{-0.62}^{+0.63}$ & $0.097_{-0.007}^{+0.008}$\\
\object{A 1413}& $0.1430$ & $6.62\pm0.14$ & $4.82_{-0.40}^{+0.44}$ & $ 7.55_{- 0.27}^{+ 0.28}$ & $50.0_{-2.1}^{+2.1}$ & $0.157_{-0.014}^{+0.016}$\\
\object{A 478}& $0.0881$ & $7.05\pm0.12$ & $7.57_{-1.02}^{+1.20}$ & $ 9.33_{- 0.43}^{+ 0.46}$ & $65.8_{-3.2}^{+3.4}$ & $0.123_{-0.017}^{+0.020}$\\
\object{PKS 0745-191}& $0.1028$ & $7.97\pm0.28$ & $7.27_{-0.70}^{+0.80}$ & $10.71_{- 0.47}^{+ 0.50}$ & $85.3_{-4.8}^{+5.0}$ & $0.147_{-0.016}^{+0.018}$\\
\object{A 2204}& $0.1523$ & $8.26\pm0.22$ & $8.39_{-0.77}^{+0.86}$ & $10.55_{- 0.39}^{+ 0.40}$ & $87.2_{-4.0}^{+4.1}$ & $0.126_{-0.012}^{+0.014}$\\
    \hline
    \end{tabular}
\end{center}
\end{table*}

\section{The data}

\subsection{The sample}
\label{sample}
The sample comprises ten nearby 
morphologically
relaxed clusters in the temperature range $[2-9]\,\keV$. We have
previously used \xmm\ data to study  the structural and
scaling properties of the total mass \citep{pap05b,app05} and of the
entropy \citep{pap06}; the  $\TX$, $\Mv$ and
$\Rv$ values derived in these papers are used in
the present Letter (Table~\ref{tab:clu}). The observations and data
reduction steps are 
fully described in \citet{pap05b}. $\Mv$ values were derived from NFW
model fits to mass profiles measured  down to 
$\delta_{\rm  obs}=600-700$,  
except for the two lowest mass clusters
($\delta_{\rm obs}\sim1400$), 
thus the $\Mv$ estimates involve some
data extrapolation. However, as discussed  in  \citet{app05}, the
$\Mv$ estimates rely solely on the physically and
observationally-motivated assumption that the best fitting NFW model
remains valid between $\delta_{\rm obs}$ and $\delta=500$, and not on
a less reliable extrapolation of density and temperature profiles.
The temperature $\TX$ was derived from a single-temperature fit to the
integrated spectrum in the $[0.1$--$0.5]\,R_{200}$ aperture,   the
inner radius defined to exclude the 
cooling core region and the outer radius chosen to ensure a 
sufficiently
precise $\TX$ estimate over the whole mass range.   This aperture corresponds to  $[0.15$--$0.75]\,\Rv$, while an aperture of
$[0.15$--$1]\,\Rv$ is used for the definition of $\TX$ in numerical
simulations and in the Chandra analysis \citep{nag07b}.  For typical decreasing temperature profiles, these $\TX\ $ values are expected to be 
slightly 
smaller  by $3$--$6\%$
  \footnote{The difference is $3\%$ for A1413 \citep[][]{app05}, a  cluster for which the temperature profiles measured up to $\Rv$  both  with \xmm\ and \chandra\ are in excellent agreement
  \citep{pa02,vik05}.
  In the simulations of \citet{nag07a}, the $[0.15$--$0.5]\,\Rv$ temperature is $6\%$ higher than that in $[0.15$--$1]\,\Rv$. A smaller difference is expected for the aperture used here.}.

\begin{table}[t]
     \caption[]{Observed scaling relations.  For each observable
       set $(B,A)$, we fitted a power law relation of the form $B
       = C(A/A_0)^\alpha$, with $A_0 = 5\,\keV;  4\times10^{13}\msol;
       2\times10^{14}\,\msol\,\keV$ for $\TX$,  $M_{\rm g,500}$ and
       $Y_{\rm X}$ respectively.   $\sigma_{\rm log,r}$ and
       $\sigma_{\rm log,i}$ are the raw and intrinsic scatter about
       the best fitting relation in the $\log$--$\log$ plane. The \MT\
       relation is the same as that given in \cite{app05}. } 
     \label{tab:rel}
     \begin{center}
    \begin{tabular}{lccccc}
    \hline
    \hline
    Relation &  $\log_{10} C $ & $\alpha$ & $\sigma_{\rm log,r}$&$\sigma_{\rm log,i}$ \\% & $\chi^2$\\
    \hline
 $h(z) M_{500}$--$\TX$   &  $14.580 \pm 0.016$ &$1.71 \pm 0.09$ & $0.064$ & $0.039$ \\%& 12.7\\
 $h(z)^{2/5} M_{500}$--$Y_{\rm X}$   &  $14.556 \pm 0.015$ &$0.548 \pm 0.027$ & $0.062$ & $0.039$ \\%& 14.4\\
$M_{500}$--$M_{\rm g,500}$  &$14.542 \pm 0.015$ &$0.803 \pm 0.040$ & $0.065$ & $0.044$\\% & 16.6  \\
 $h(z) M_{g,500}$--$\TX$   &  $13.651 \pm 0.010$ &$2.10 \pm 0.05$ & $0.048$ & $0.036$ \\%& 12.7\\
$h(z)^{2/5} M_{\rm g,500}$--$Y_{\rm X}$  &   $13.619 \pm 0.008$ &$0.678 \pm 0.014$ & $0.017$ & - \\%& 5.7 \\
$f_{\rm g,500}$--$Y_{\rm X}$  &   $-0.939 \pm 0.016$ &$0.133 \pm 0.028$ & $0.067$ & 0.044 \\%& 5.7 \\
     \hline
    \end{tabular}
 \end{center}
\end{table}

 \begin{figure}[t]
\centering
\includegraphics[ width=0.85\columnwidth, keepaspectratio]{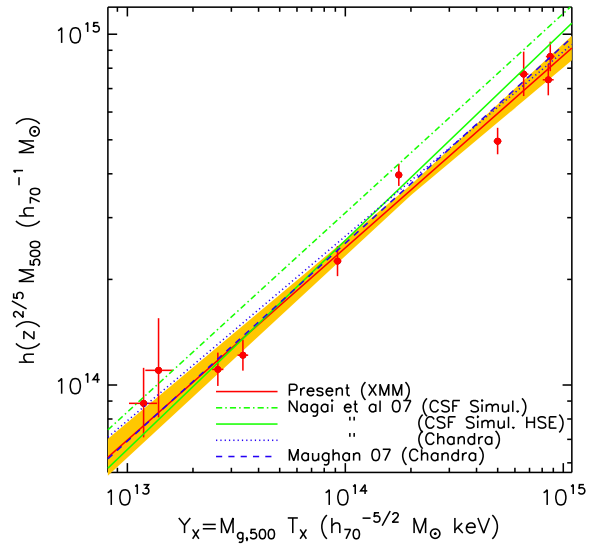} 
 \caption{\label{my} The \MY\ relation as seen by \xmm\ from a sample
   of 10 local relaxed clusters. The red solid line is the best fitting
   power law and  the shaded orange area corresponds to the $1\,\sigma$
   uncertainty. The predicted relation from numerical simulations
   including cooling and galaxy feedback \citep{nag07b} is over-plotted
   as a green dot-dashed line (true mass) and as a green solid
   line (mass estimated from mock X-ray observations and the HE
   equation). The dotted and dashed blue lines are the observed
   relations derived from Chandra data by \citet{nag07b} and
   \citet{mau07} respectively (see text). } 
\end{figure}

%===============================================================
\begin{figure*}[th]
\begin{center}
\begin{minipage}[t]{0.9\hsize}
\resizebox{\hsize}{!} {
\includegraphics{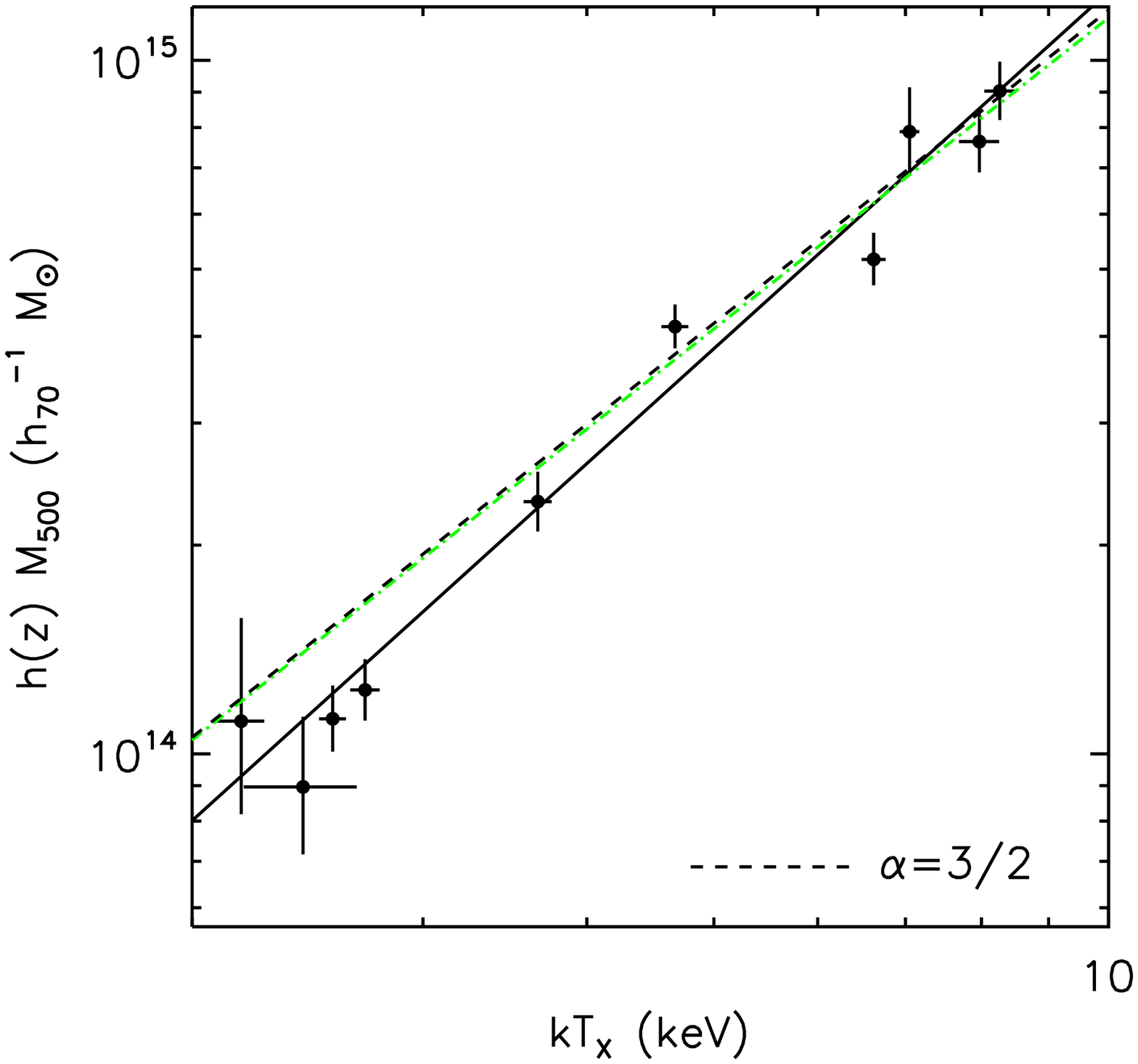}
\hspace{8mm}
\includegraphics{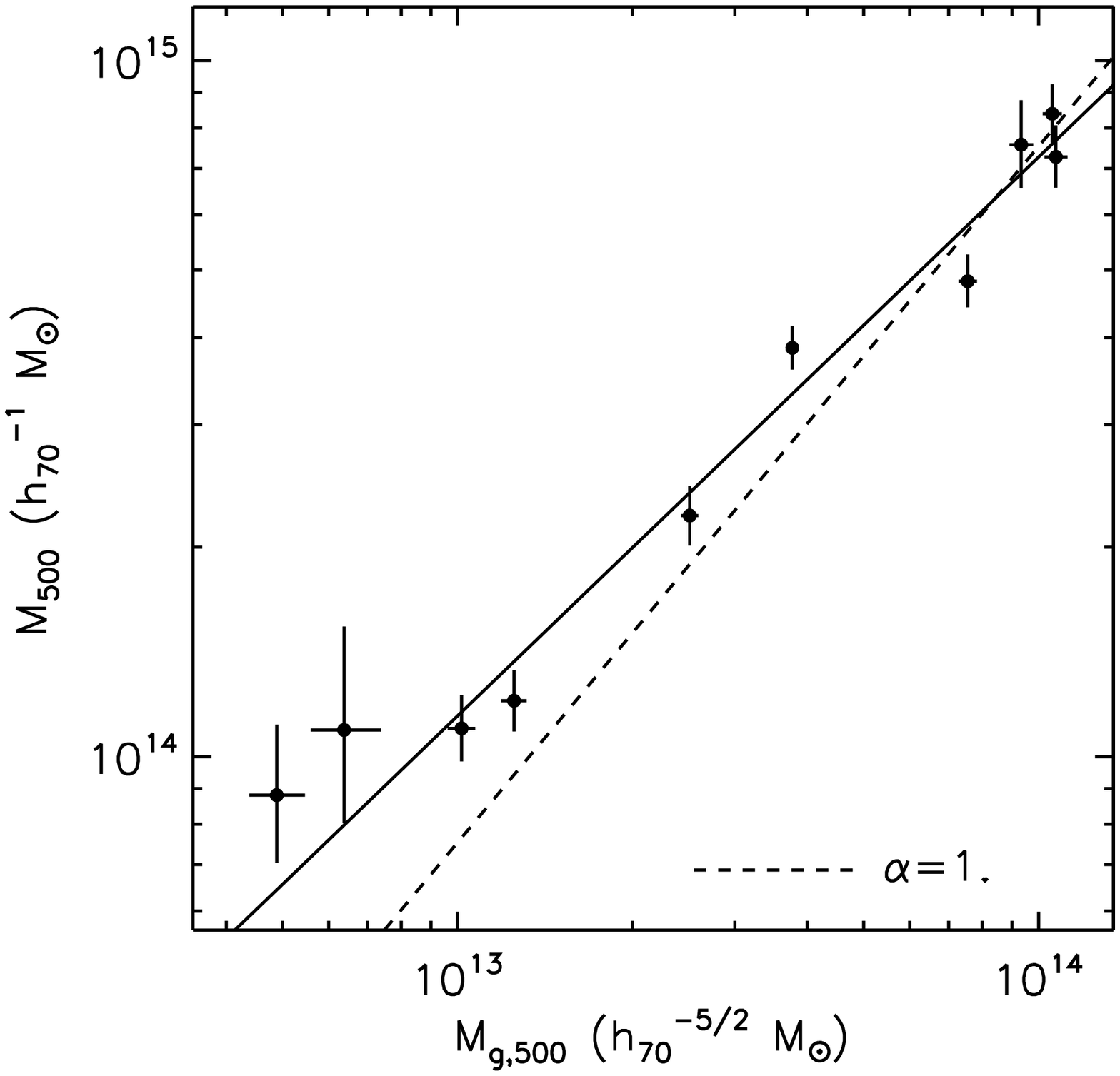}
\hspace{8mm}
\includegraphics{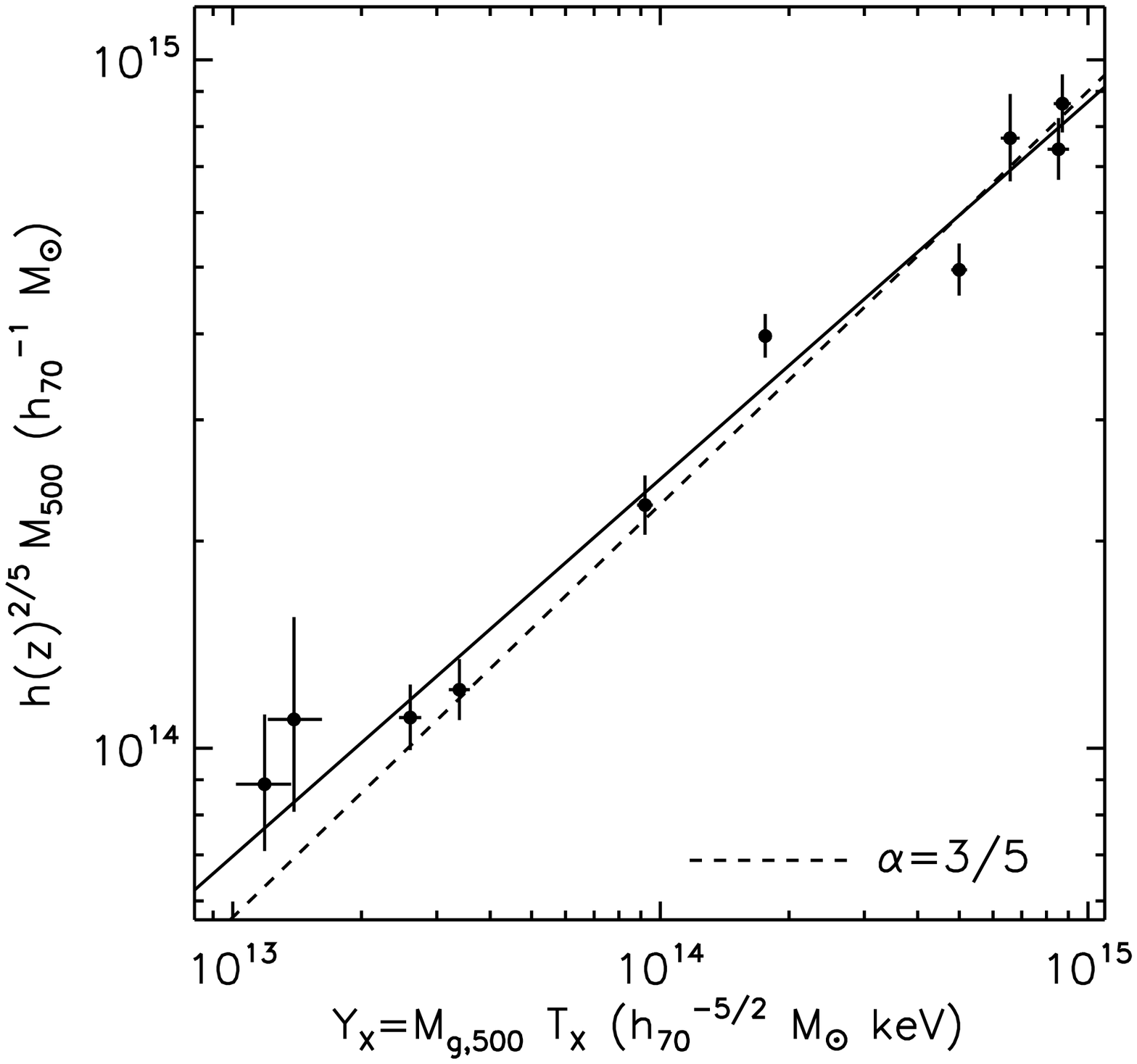}
} 
\end{minipage}\\[2mm]
\begin{minipage}[t]{0.9\hsize}
\resizebox{\hsize}{!} {
\includegraphics{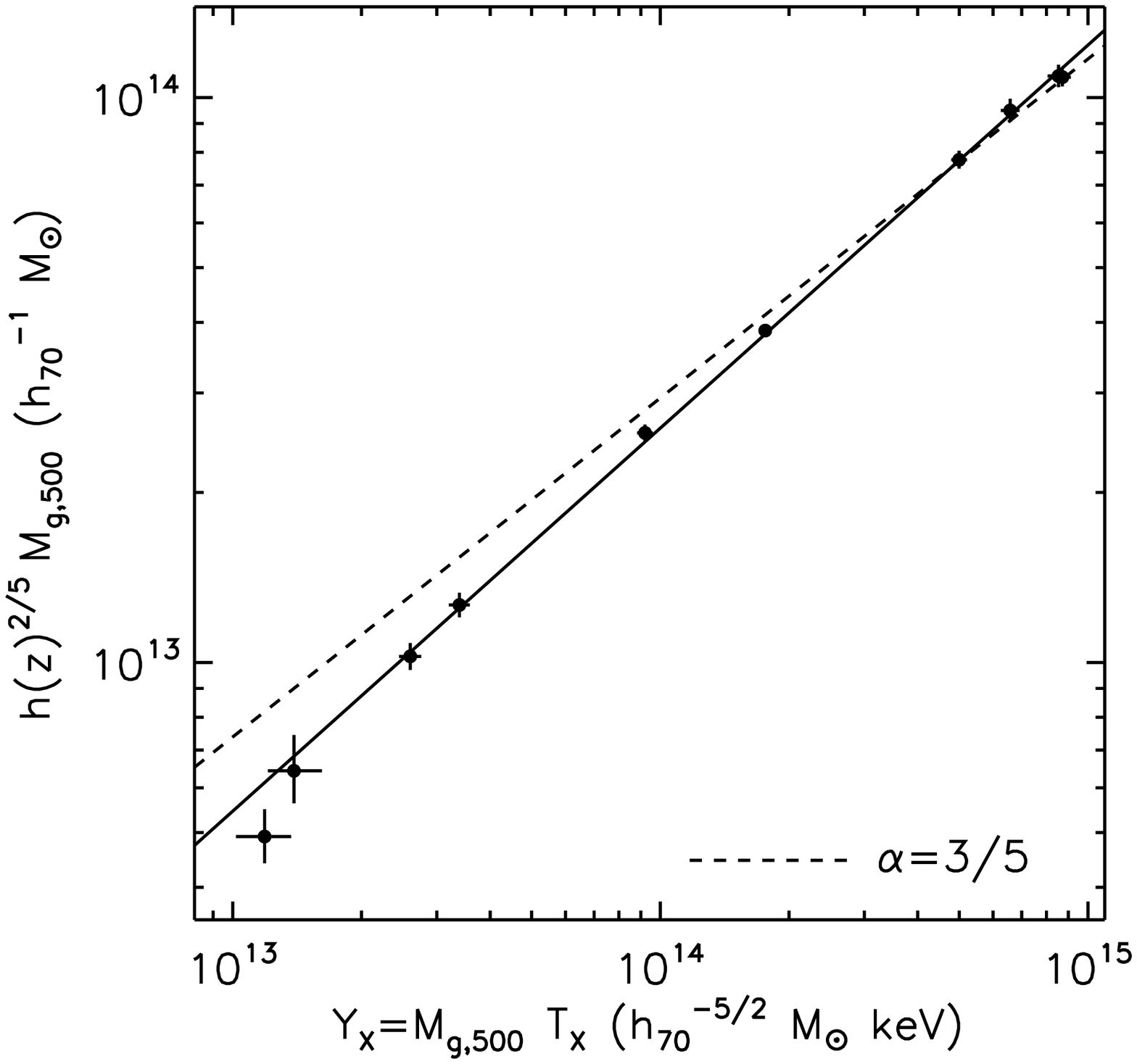}
\hspace{8mm}
\includegraphics{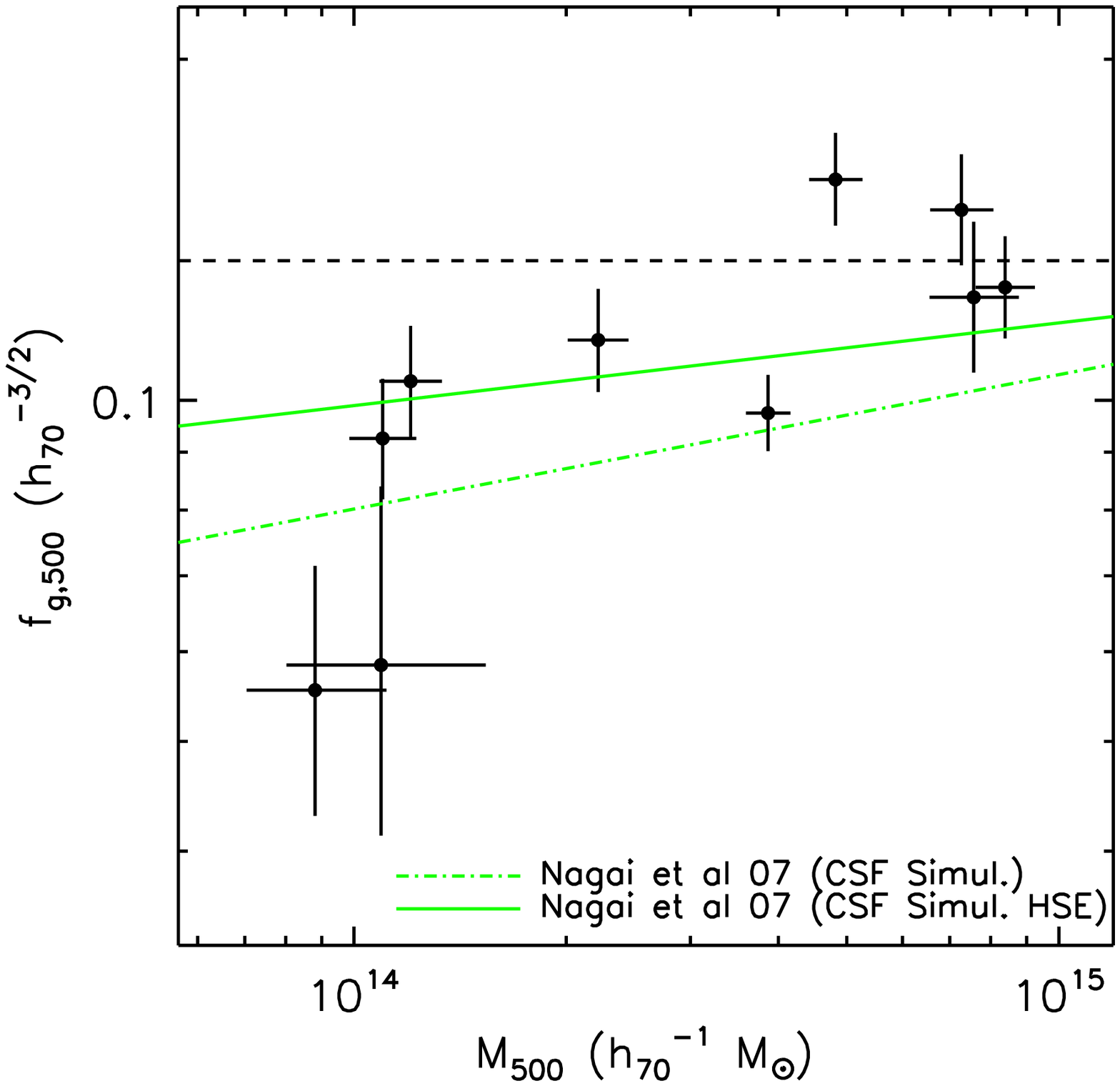}
\hspace{8mm}
\includegraphics{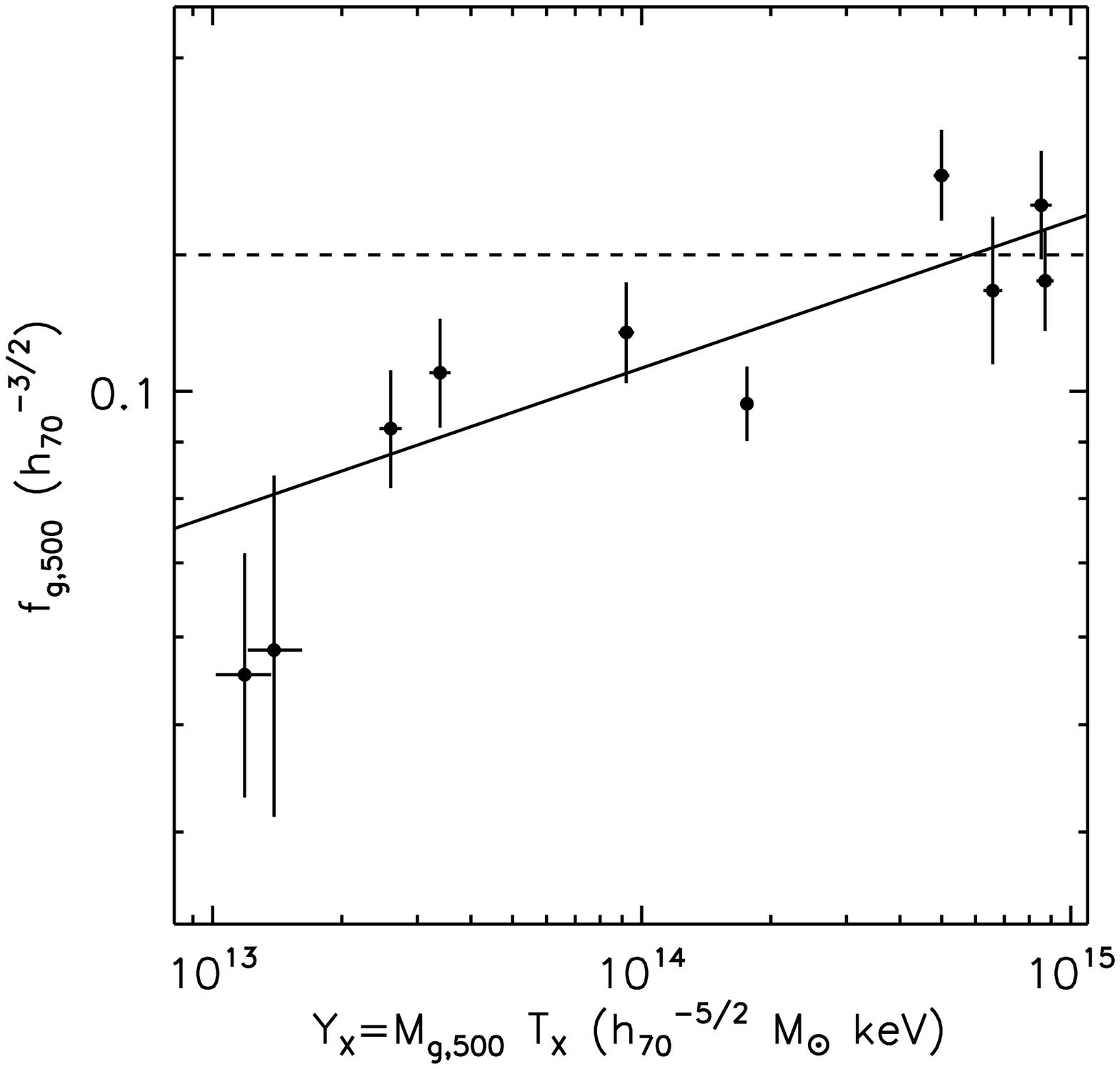}
} 
\end{minipage}
\caption{\label{rel}   Correlations between X-ray
  observables. Solid lines: best fitting power law relations.  Dashed lines: standard self-similar
  relation (slope indicated in each figure) normalised to data from the three
  most massive clusters. Green
  dotted line in top-left panel:  best fitting power law  \MT\  relation
  for the hot cluster sub-sample \citep{app05}.  Green  lines in bottom-middle
  panel: gas mass fraction from the \MMg\ relation in the numerical simulations of \citet{nag07b},  using the true  mass (dash-dotted line) and the HE mass (full line). } 
\end{center}
\end{figure*}
%===============================================================

The integrated gas mass depends sensitively on the gas density at
large radius.  To compute $\Mgv$, we re-derived the gas density
profile from the emissivity corrected surface brightness profiles
using the deprojection and PSF-deconvolution technique recently
developed by \citet{cro06}. This derivation is free of any
assumption on profile shape, such as power law behaviour at large
radius, a feature common to all analytical fitting models used thus far
\citep[e.g.][]{pa02,pap05b,vik06}. Furthermore, the statistical errors
are readily estimated from a built-in Monte-Carlo procedure
\citep[see][for details]{cro06}. For the present sample, there is
excellent agreement  between the deprojected density profiles and the 
analytical model profiles derived in our previous work
\citep{pap05b,pap06}. The significant differences are in the very
central regions of some clusters  \citep[e.g.][Fig 12]{cro06} and for
A2597 at large radii, where the deprojected profile is slightly
steeper than the model profile. The gas mass estimated
with the two methods differs by less than $3\%$, except for  A2597
($8\%$ difference). For all clusters, except for A1983 and MKW9, the
surface brightness profiles extend at least up to $\Rv$, or very close to
it, so that extrapolation uncertainty is not an issue. For A1983 and MKW9, the $\Mgv$
estimated from extrapolation in the $\log$--$\log$ plane are $31\%$
(A1983) and $67\%$ (MKW9) larger than the gas mass measured at
$\delta_{\rm obs}$
; as shown below, these points do not however have a
significant effect on the results. 

The  resulting  $\Mgv$, $\fgv$ and $\YX$ values are listed in
Table~\ref{tab:clu}. Errors  on $\Mgv$ include both statistical errors
and errors due to uncertainties in $\Rv$, which are summed
quadratically, with the latter dominating the error budget. 

\subsection{Scaling relations}

For each observable set $(B,A)$ we fitted a power law relation of the form
$h(z)^nB = C(A/A_0)^\alpha$, where 
$h(z)$  is the Hubble constant
normalised to its present value 
and $n$ is fixed to the expected
scaling with $z$.  The fit was performed using linear
regression in the $\log$--$ \log$ plane, taking into account  the
errors on both variables
\citep[\texttt{FITEXY};][]{numrec}.  The pivot
point $A_{0}$ is chosen so that the normalisation and slope are nearly
independent parameters. For the $M_{500}$--$Y_{X}$ relation for
instance, the covariance in $\log(C)$ and $\alpha$ normalised to the
product of their standard errors is 0.042. The resulting values are
given in Table~\ref{tab:rel}, and the various correlations are plotted
in Fig.~\ref{my} and   Fig.~\ref{rel}. 
Table~\ref{tab:rel} also shows the raw and intrinsic scatter about 
the best fitting  relations in the  $\log$--$ \log$ plane.
The raw scatter was estimated using the vertical distances to the  
regression line, weighted by the error. The  
intrinsic scatter was computed from the quadratic difference between the  
raw scatter and the scatter expected from the statistical errors.

The regression method is strictly valid only if the intrinsic
scatter is negligible as compared to the statistical scatter; in fact
they are of the same order (Table~\ref{tab:rel}). We verified that the
results are unchanged using the variation of the method discussed in
\citet{pap06}. 
Finally, the \MY\ relation is robust to exclusion of
A1983 and MKW9, for which data extrapolations were required (see above): the differences are at the $+0.4\,\sigma$  and $-0.2\,\sigma$ levels for the slope and normalisation, respectively.

\section{Discussion and conclusions}
\subsection{Comparison with theoretical predictions}
The slope of the observed relation:
\begin{equation}
h(z)^{2/5}\Mv = 10^{14.556 \pm 0.015} \left[\frac{\YX}{2\times10^{14}\,{\msol}\,\keV}\right]^{0.548 \pm 0.027}{\rm h_{70}^{-1}\,\msol}
\end{equation}
is slightly smaller than the standard self-similar  value
($\alpha=3/5$),  at the $1.9\,\sigma$ significance level, consistent
with the  \MMg\ and \MT\  relations  (Table \ref{tab:rel} and
Fig.~\ref{rel} top panel). The  \MMg\ relation is shallower than
expected, reflecting the increase in gas mass fraction with mass
(Fig.~\ref{rel} bottom-middle panel), while the  \MT\ is steeper.
At a given mass the gas mass is smaller and the temperature is
higher, leading to a partial cancellation in the product $\YX= \Mg
\TX$ (see also below).    

The observed normalisation is $\sim 20\%$ smaller over the whole $\YX$ range
than that derived from numerical simulations including cooling and galaxy feedback \citep{nag07b}, while the observed slope is
consistent with the predicted slope, $\alpha = 0.568\pm 0.006$, within
the $1\,\sigma$ error (Fig.~\ref{my}).  
Better agreement is obtained
with the  simulated $M_{500}^{HE}$--$\YX$ relation,  where $M_{500}^{HE}$ is the mass estimated from mock X--ray
observations and the HE equation.  Although
the predicted slope, $\alpha = 0.596\pm 0.010$,
%(close to the self-similar value), 
is slightly higher,  the difference in normalisation drops
to  $\sim\,8\%$ ($2.4\,\sigma$) at $\YX = 2\times10^{14}\,{\msol}\,\keV$.   As
discussed by \citet{nag07b}, the offset in normalization, also observed
with \chandra\ data,  may arise from an underestimate of the true mass
by the HE equation, perhaps due to residual non-thermal pressure
support.  These numerical simulations also
predict a hot gas mass fraction systematically smaller than observed
(Fig.~\ref{rel} bottom-middle panel). The difference is smaller for simulated  $\fgv$ using $M_{500}^{HE}$ and again could be due, in part, to  biases in X--ray mass estimates. Nevertheless, there may also be an underestimate of $\fgv$ in the simulations, possibly due in part  to  over-condensation of hot gas into the cold dense phase \citep[][]{nag07b}. This would  contribute to the offset, by shifting the \MY\ relation to the left in the $\log$--$\log$ plane. Finally, as the normalization  depends on $\TX^{0.6}$,  the difference in the exact definition of $\TX$ (see Sec.~\ref{sample}) could contribute by $\lesssim 4\%$ to the offset.
  
\subsection{Comparison with \chandra\ results}

Our \MY\ relation is very similar to that derived by \citet{nag07b}
from the \chandra\ data presented in \citet[][see our
Fig.~\ref{my}]{vik06}. The slope $\alpha =0.526  
\pm0.038$ is consistent with our value, $\alpha = 0.548\pm0.027$,
and the normalisation at $\YX = 2\times10^{14}\,\msol\,\keV$, $\Mv =
3.82\times10^{14}\,{\rm h_{70}^{-1}}\,\msol$, is higher than our value,
$(3.60\pm0.13)\times10^{14}\msol$, at only the $1.6\,\sigma$ level. Even
better agreement is obtained with the best fitting relation quoted  by
\citet[][dashed line in Fig.~\ref{my}]{mau07}, derived from the same
data excluding the lowest mass cluster
(A. Vikhlinin, priv. communication). Here the  slope ($\alpha
=0.564$) is closer to the self-similar value, as we have found, and
the difference in normalisation is less than $5\%$ over the whole mass
range.

\subsection{Comparison of  mass proxies for relaxed clusters}

For {\it relaxed} clusters, \citet{kra06} found similar scatter in the
\MT\  and \MMg\ relations ($\sigma_{\rm log} = 0.055$ and 0.047
respectively), but two times less scatter in the \MY\ relation (0.022).
We can compare with the present data, the statistical quality allowing us to estimate the intrinsic scatter for the
first time. %%In fact 
The scatter 
(Table~\ref{tab:rel}) 
is the same for the \MY\ and \MT\  relations
($\sigma_{\rm log,i} = 0.039$) and slightly larger for the \MMg\
relation ($\sigma_{\rm log,i} = 0.044$). The latter may reflect that
the \MMg\ relation is not actually a power law: 
the gas mass fraction appears constant at $\Mv
\gtrsim 2-3\times10^{14} \msol$, with a progressive drop at lower mass (Fig.~\ref{rel} bottom-middle panel). 

In fact the behaviour of $\fgv$ appears to be the primary factor
driving the scatter in the \MY\ relation. The \MgY\ relation is
extremely tight (Fig.~\ref{rel}
and Table~\ref{tab:rel}), being well fitted by a power law 
%($\chi^2_{red}=0.7$),
with no measurable scatter, in spite of the precision of the
data. Since $\Mv=\Mgv /\fgv$, the scatter in the \MY\ relation simply
reflects the scatter in the $\fgv$--$\YX$ relation (cf. top and
bottom left panels of Fig.~\ref{rel}). This scatter could arise from  true scatter in $\fgv$ and/or scatter in the X--ray mass  to true mass ratio, e.g., due to variations in the magnitude of nonthermal pressure support.
Note that a low-scatter
correlation between $\Mgv$ and $\YX$ is expected: it is
straightforward to show that the logarithmic scatter in the \MgY\
relation is 1/3 of the scatter in the \MgT\ relation for
$\Mgv \propto \TX^{\sim 2}$ (Table~\ref{tab:rel}).

In terms of observed scatter in the relation with mass,  $\YX$
thus does not appear to be a better proxy than $\TX$, and is only slightly
better than $\Mgv$.  However we caution against over-inerpretation. Firstly, the present results are for relaxed
clusters only: with the current data we cannot check if the scatter is
insensitive to dynamical state \citep{kra06,poo07}. Secondly, the 
scatter estimates should be
confirmed using larger cluster samples with stricter
selection criteria.

However, in terms of functional dependence with mass, $\YX$ is clearly
a better proxy than $\Mgv$: it is better fitted by  a simple
power-law, and has a slope closer to the standard self-similar value
(Table~\ref{tab:rel}).  Furthermore, although the  quality of the
power law fits to \MT\ and \MY\ are formally similar ($\chi^2/{\rm
  d.o.f}\sim13/8$), with similar ($\sim2\,\sigma$) deviations from the standard slope, there is some indication that $\YX$ is
also a better proxy than $\TX$ in this regard. The slope of the \MT\
relation may depend on mass range \citep{app05}, reaching the standard value  when cool clusters are excluded,  but the slope of the \MY\ relation remains stable in that case ($0.7\,\sigma$ difference). 

\subsection{Concluding remarks}

Our results suggest that the various mass scaling relations
might be better understood by considering the gas thermal energy
($\YX$) and mass ($\fgv$) as its most fundamental properties.  
Let us suppose that the thermal energy content of the gas is the quantity most closely related to the mass (i.e. the best mass proxy is indeed $\YX$), and
that its relation with mass has a quasi-standard slope. Let us further note that the gas mass fraction appears constant at high
mass, with a progressive decrease below 
a 'break' mass (reflecting gas loss or incomplete accretion in low
mass systems due to non gravitational effects). Since
$M/T^{3/2}$ varies as $(M/\YX^{3/5})^{5/2} f_{\rm g}^{3/2}$, one
then expects a steepening of the \MT\ relation at low mass, with a
standard slope at high mass.

A deeper understanding of the mass scaling relations will come from
the X-ray study of larger unbiased samples of local clusters,
such as REXCESS \citep{boh07}, combined with lensing data.
This is necessary
to ascertain the dependence of the \MY\ relation on dynamical
state, and to calibrate its normalisation and slope. This step is essential because the use of $\YX$ as a mass proxy, as in the case of $\TX$, requires a detailed understanding of non-gravitational effects, in particular of the impact of cooling and feedback on the fraction of primordial gas that remains in the gravitationally bound hot phase. Precise measurements at $z=0$ are
needed to constrain models, on which one must rely for high z
studies.  
Significant progress is also expected from forthcoming SZ data (e.g from the Planck Surveyor all sky survey),
especially if combined with \xmm\ or Chandra data, which will allow a
full study of the $M$--$Y_{\rm SZ}$ relation.

\begin{acknowledgements}
    We thank A. Kravtsov and D. Nagai  for useful comments on the manuscript, and the referee for a speedy and pertinent response. 
\end{acknowledgements}

\end{document}